\documentclass[journal,twoside,web]{ieeecolor}
\usepackage{generic}
\usepackage{cite}
\usepackage{amsmath,amssymb,amsfonts}
\usepackage{algorithmic}
\usepackage{graphicx}
\usepackage{textcomp}
\newcommand{\AlBScN}{Al$_{0.71}$B$_{0.06}$Sc$_{0.23}$N}
\newcommand{\AlScN}{Al$_{0.85}$Sc$_{0.15}$N}
\newcommand{\AlBScNs}{Al$_{0.71}$B$_{0.06}$Sc$_{0.23}$N }
\newcommand{\AlScNs}{Al$_{0.85}$Sc$_{0.15}$N }
\def\BibTeX{{\rm B\kern-.05em{\sc i\kern-.025em b}\kern-.08em
    T\kern-.1667em\lower.7ex\hbox{E}\kern-.125emX}}
\begin{document}
\title{Three Million Years Opposite State Data Retention in Partially Switched 
Wurtzite Ferroelectrics}
\author{Maike Gremmel, Roberto Guido, \IEEEmembership{Member, IEEE}, Victor Witte, Thomas Mikolajick, \IEEEmembership{Fellow, IEEE}, Uwe Schroeder, \IEEEmembership{Senior Member, IEEE}, Simon Fichtner, \IEEEmembership{Member, IEEE}
\thanks{This work was supported by the German Research Foundation under Grant 458372836 and 458372836. S. Fichtner was funded by the European Union, the European Chips Joint Undertaking (CHIPS JU) and its members under the FIXIT project (Grant Agreement No. 101135398). Views and opinions expressed are however those of the author(s) only and do not necessarily reflect those of the European Union. Neither the European Union nor the CHIPS JU or its members can be held responsible for them.
T. Mikolajick and U. Schroeder were supported by the Saxonian State budget approved by the delegates of the Saxon State Parliament.}
\thanks{M. Gremmel, V. Witte, and S. Fichtner are with the University of Kiel, Department of Material Science, Kaiserstraße 2, 24143 Kiel, Germany (e-mail: magr@tf.uni-kiel.de, sif@tf.uni-kiel.de)}
\thanks{R. Guido, T. Mikolajick, and U. Schroeder are with NaMLab gGmbH, Noethnizer Strasse 64a, 01187 Dresden, Germany.}
\thanks{S. Fichtner is also with the Fraunhofer Institute for Silicon Technology, Fraunhoferstrasse 1, 25524 Itzehoe, Germany and T. Mikolajick is also with TU Dresden, Noethnitzer Str. 64, 01187 Dresden,Germany}}

\maketitle

\begin{abstract}
Ferroelectric memories based on wurtzite-structured compounds are projected to store information for more than three million years at 150$^\circ$C. These results are extracted by combining switching kinetics with the nearby-electrode charge injection model for opposite state retention in ferroelectric random access memory. This impressive performance is greatly aided by switching only a fraction of the total polarization to store data, in order to limit the initial imprint (i.e. coercive field) variation of the devices - an effect that is confirmed for both \AlScNs and \AlBScN. Paradoxically, yet systematically, this reduction in initial imprint consistently results in larger switching polarization after a given time, compared to the fully switching state and 4 to 7 orders of magnitude improvement in predicted opposite state retention. In addition, partial switching is able to simultaneously boost switching endurance, making it a promising strategy for improved reliability of ferroelectric devices with large spontaneous polarization, particularly wurtzite-structured compounds. 
\end{abstract}


\section{Introduction}
\label{sec:introduction}
\IEEEPARstart{D}{ata} retention in Ferroelectric Random Access Memory (FeRAM) is typically determined by opposite state retention (OSR) (i.e. the ability to perform switching read-operations with a given voltage after a given time), rather than gradual depolarization (same state retention, SSR). \cite{Silva2023RoadmapDevices,Alcala2022TheHfO2Capacitors,Vishnumurthy2024Ferroelectricinvited,Alcala2023TheCapacitors,Sunbul2023AAnnealing} OSR itself is determined by imprint, i.e. differences in the absolute value of the positive and negative coercive field $E_c$. Once the negative or positive $E_c$ approaches or surpasses the boundaries of the read-window that is defined by the available supply voltage, part of the polarization available to the read-operation will be lost. Typically, imprint continues to increase over time.\cite{Tagantsev_Imprint_2004} This leads to a memory loss once the remaining ferroelectric switching charge inside the read-window falls below the threshold given by the read-circuit for determining the state of the memory cell. In the following text a threshold of 26.5 $\mu$C/cm$^2$ is used.\cite{IEEE_IRDS_2022} 

Due to their large remanent polarization ($P_r$), and excellent thermal stability, the emerging wurtzite ferroelectrics possess characteristics that can make them the new cutting-edge long retention/high temperature non-volatile memory compounds. \cite{Fichtner2019AlScN:Ferroelectric,Islam2021OnFilms} In addition, their large $E_c$  can be seen as a benefit, as it decreases the ratio between imprint field and $E_c$. In this context, we recently showed that imprint relative to $E_c$ stays below 10\% for wurtzite \AlScNs after 10$^6$ s of accelerated aging at 150$^\circ$C. This is significantly less than what is typically observed in e.g. HfO$_2$-based ferroelectrics. \cite{Guido2025FerroelectricDynamics,Vishnumurthy_Imprint_2024,Barbot_Imprint_2025} At the same time, the variation of imprint over time in wurtzite \AlScNs was observed to follow the well established nearby-electrode charge injection model developed by Tagantsev et al.\cite{Tagantsev_Imprint_2004,Guido2025FerroelectricDynamics}

\begin{figure*}[h]
	\centering
	\includegraphics[width=\textwidth]{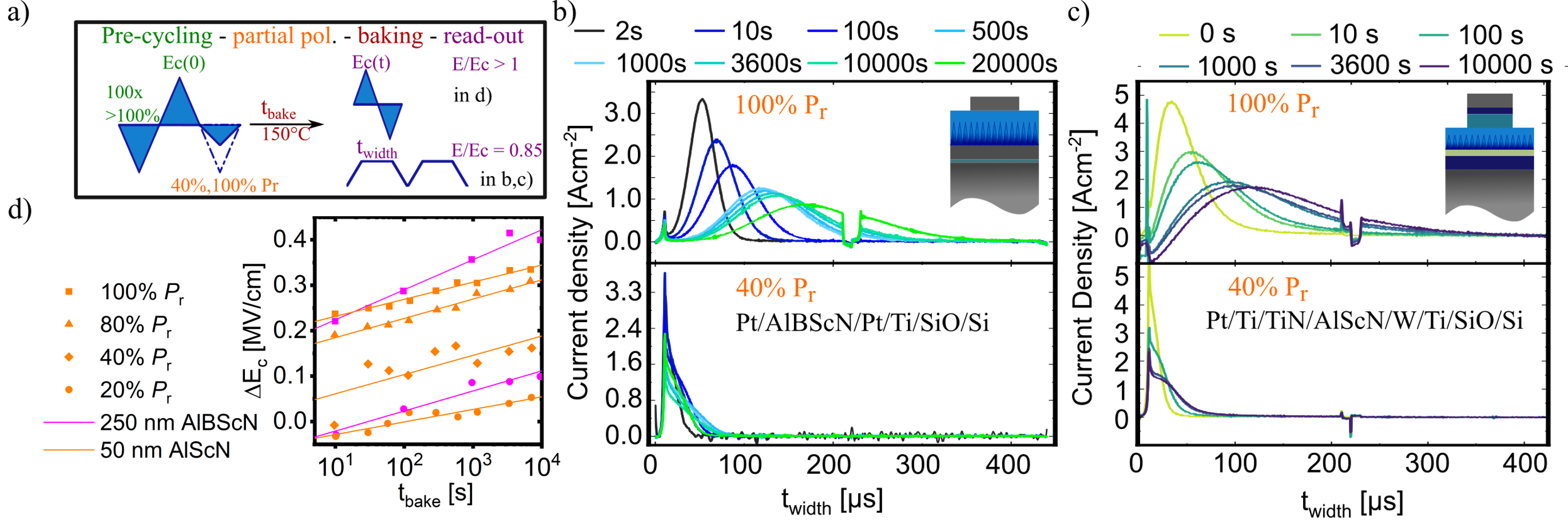}
	\caption{a) Electrical characterization sequence for OSR. b) Displacement current density for fully and partially switched (40\% $P_r$) \AlBScNs (270 nm) and c) \AlScNs (60 nm) after different accelerated aging times $t_\text{bake}$ for the dual trapezoidal waveform in panel a). d) Imprint over $t_\text{bake}$ for \AlBScNs and \AlScNs (50 nm) for fully and partially switched states.}
	\label{fig1}
\end{figure*}

In addition to this variation, changes in $E_c$ can occur as more immediate consequence of switching.\cite{Gremmel2024TheAl0.70Sc0.30N} For wurtzite ferroelectrics, this \emph{abrupt} variation can easily account for more than 50\% of the total imprint [Figure \ref{fig1} d)]. Therefore, tailoring the switching event itself can be a powerful lever to improve OSR. Based on 270 nm thick \AlBScNs as well as 50 to 60 nm thick \AlScNs capacitors, we demonstrate in the following that imprint can be largely compensated by using partial switching. This approach is able to boost projected OSR by an impressive 4 to 7 orders of magnitude towards quasi-perpetual storage at 150°C. Therefore, wurtzite ferroelectrics allow to \emph{trade} part of their exceptionally large $P_r$ for superior data protection - while at the same time also offering significant advantages to overall switching endurance.\cite{Cho2026WriteAlScN} 

\section{Results}
\subsection{Opposite State Retention}
In order to allow a reliable projection of OSR beyond the measured maximum accelerated aging time $t_\text{bake}$, two datasets are required: $\Delta E_c(t_\text{bake})$ for fully and partially switched states as well as the switching polarization $P_\text{sw}$ over time $t$ for different $E/E_c$, as in a regular switching kinetics study. Their combination allows to project $P_\text{sw}(t_\text{bake})$.  The pulse sequence used for the extraction of $\Delta E_c(t_\text{bake})$ is displayed in Figure \ref{fig1} a), as is the scheme for the analysis of the {read}-operation. The latter consist of a trapezoidal read- and write-pulse.  

Figure \ref{fig1} d) shows the evolution of the imprint-induced $\Delta E_c(t_\text{bake})$ = $E_c(t_\text{bake}) -  E_c(0)$. With increasing t$_\text{bake}$ at 150 °C, $\Delta E_c$ increases logarithmically for all investigated states, in accordance with the charge injection model by Tagantsev et al.\cite{Tagantsev_Imprint_2004} Imprint after $10^4$ s stays relatively modest at around $0.1E_c$. As all states increase with broadly similar slope, the underlying mechanism is expected to be comparable - likely thermally activated charge injection and trapping at the electrode/ferroelectric interfaces. However, a major difference between the states lies in the onset of imprint, reflected by a progressively lower intercept with the $\Delta E_c$ axis for lower $P_\text{sw}$. Consequently, Figure \ref{fig1} b), c) confirms that the risk of \emph{losing} switching charge outside of the read-pulse is significantly smaller for $P_\text{sw} = 40\%$, as the current response is localized at the very beginning of the read-event. On the contrary, for $P_\text{sw} = 100\%$, switching charge transitions from the read-pulse to the control-pulse already after 500 s at $t_\text{bake}$.
Figure \ref{fig1} b), c) furthermore confirms that an overall reduction of $E_c$ is responsible for the lower imprint, as the 40\% $P_r$ read-operation does not result in a mere sub-loop of the 100\% $P_r$ read-operation.  
This reduced E$_c$ can be attributed to an enhanced presence of mobile domain walls or less charge injection at lower applied fields.\cite{Gremmel2024TheAl0.70Sc0.30N} The result is a larger $E/E_c$, which promotes faster switching.
In order to quantify the OSR improvement due to partial switching, the $E_c$ dependent switching kinetics of the samples were evaluated by applying the Kolmogorov–Avrami–Ishibashi (KAI) model. While the \AlScNs film could be modeled with an Avrami exponent of $n=2$, \AlBScNs required $n=4$, reflecting different transition regions towards the simultaneous non-linear nucleation and growth model.\cite{Yazawa2023AnomalouslyModel} This is, however, of little consequence for the quantitative analysis in this study, since the KAI fits reflect our measured curves sufficiently well for all evaluated values of $E/E_c$. In addition, partially switched films exhibit a characteristic switching time $t_0$ that is reduced by a factor of 2 compared to fully switched films. Together with the larger $E/E_c$, this reduction of $t_0$ further contributes to faster switching and thus longer OSR due to stronger confinement of the switching current inside the read-pulse.\cite{Guido2025FerroelectricDynamics} 

\begin{figure*}[h]
	\centering
	\includegraphics[width=\textwidth]{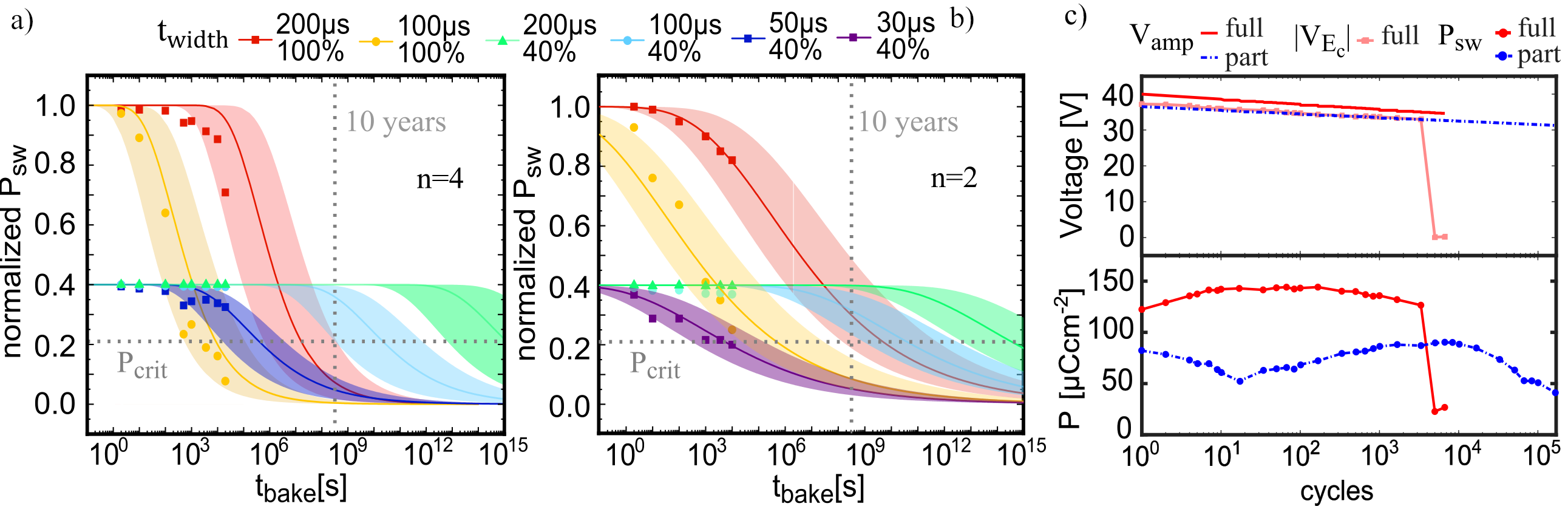}
	\caption{ a) Measured and projected OSR for 270 nm thick \AlBScNs and b) 60 nm thick \AlScN, for both fully (100\%) and partially (40\%) switched states. The $26.5$ $\mu$C/cm$^2$ limit for successful readout ($P_\text{crit}$) as well as 10 years OSR are marked by dotted lines.  c)   Endurance measurement for fully and partially switched capacitors (60 nm \AlScN).}
	\label{fig2}
\end{figure*}

Figure \ref{fig2} a), b) displays the measured and projected $P_\text{sw}(t_\text{bake})$ for 270 nm thick \AlBScNs and 60 nm thick \AlScN.  Overall, the projected OSR agrees well with the experimental data within the range of measurement uncertainty. 

Most notably, for both capacitor stacks - \AlBScNs in Figure \ref{fig2} (a) and \AlScNs in Figure \ref{fig2} (b) - the $P_\text{sw}$ of partially switched states begins to exceed that of fully switched films already within in the directly measured interval of up to $10^4$ s, when integrating (\emph{reading}) for 200 $\mu$s. This leads to the paradoxical situation where the read-operation for the partially switched capacitor results in a larger charge response compared to fully switched devices, even though the initial $P_\text{sw}$ of the latter was more than twice as large as that of the partially switched devices. 

Going beyond the directly probed interval of $10^4$~s, the projected OSR confirms the generality of this observation for different read-times. For the \AlScNs sample (Figure \ref{fig2} b), OSR improves by 4 to 5 orders of magnitude, whereas the increase is even 6 to 7 orders of magnitude for the \AlBScNs sample. In both cases, the expected OSR reaches more than three million years at 150$ ^\circ$C for 200 $\mu$s read-time. 

%However, for \AlBScN an adjustment of the effective Avrami exponent from n = 2 to n = 4 impacts the overall retention time of the samples. In fully switched \AlBScN, the onset of retention loss starts at higher $t_{bake}$ compared to \AlScN.  However due to the higher $n$, the slope is steeper and the critical polarization value for devices is reached earlier.

% Shown in \ref{fig2}(a) is the polarization over a number of cycles for both a fixed Voltage and an adjusted Voltage. Looking at the number of cycles after which breakdown occurs, an improvement from 10$^3$ to 10$^5$ can be achieved by a (stepwise) reduction of voltage while keeping the full P$_{sw}$. The Partial polarization state can be kept constant through an adjusted Voltage application  and leads to an noticable improvement of endurance. These results are coherent to observations by Cho et al, who also reported improved endurance by partial switching on a similar setup for \AlScN. 
% \textit{$->$	Even though writing with a fixed electric field may change the percentage of switched volume, the overall lifetime still exceeds the one of fully switched films $->$ adjustment of E necessary to stay at 40 \%}
\subsection{Switching Endurance}
\color{black}
In addition to characterizing OSR, we verified that partial switching does indeed result in the expected endurance increase with respect to the maximum numbers of switching cycles.~\cite{Cho2026WriteAlScN} Figure~\ref{fig2}(a) shows the evolution of $P_\text{sw}(n)$ as a function of cycle number $n$ for both full-switching and partial-switching operation. To maintain partial switching, the applied voltage has to be continuously adapted. In addition to the imprint-induced shift $\Delta E_{\mathrm{c}}$, a cycle-dependent contribution of \mbox{$\Delta E_{\mathrm{c}}(n) = -1.4 \cdot \log(n)$} was observed during switching from negative to positive $P$.

To maintain partial switching, the applied voltage amplitude was adjusted accordingly. Assuming that partial switching follows a similar dependence, $\Delta E_{\mathrm{c}}(n) \propto -\log(n)$ and that the distribution of switched domains remain invariant, a sufficiently stable partially switched state corresponding to $34.7\%$ to $62.5\%$ of the full polarization was maintained up to $10^{5}$ cycles. Beyond this point, the polarization deviated from the targeted behavior more strongly, and a hard breakdown occurred after approximately $2.8 \cdot 10^{5}$ cycles.

Compared to full switching, the breakdown endurance therefore improved %from $10^{3}$ to $10^{5}$
by two orders of magnitude. This confirms that maintaining a constant partial polarization state is feasible via an appropriate $V_{\mathrm{amp}}(n)$ and significantly enhances device endurance. Therefore, partial switching can be considered highly desirable from two central device metrics: data retention and switching endurance.

\section{Methods}

All measurements were done on co-sputtered films of 270 nm thick \AlBScNs and 50 to 60 nm thick \AlScN.  The sample preparation methods for all films have previously been reported.\cite{Guido2025FerroelectricDynamics, Gremmel_boron_2025, Schonweger2023InGrainV}
Thermal treatment was performed on a hot plate at 150$^\circ$C in ambient air. Electrical characterization was conducted at room temperature using an aixACCT TF Analyzer 3000. The specific treatment schemes are displayed in Figure \ref{fig1} a).
The methodology for projecting $P_\text{sw}(t_\text{bake})$ beyond the measured time has previously been described.\cite{Guido2025FerroelectricDynamics}

\section{Conclusion}
Our findings highlight the substantial potential of wurtzite ferroelectrics for quasi-perpetual data storage. In contrast to alternative approaches for extreme data retention, such as laser-written memory crystals,\cite{Allison2026} wurtzite ferroelectrics have the distinct advantage that electrical read- and {(re-)write-procedures} can be employed. 

The observed mechanism for OSR prolongation through partial switching can likely be extended to other ferroelectrics were OSR is limited by charge injection. However, wurtzite ferroelectrics benefit in particular from this approach due to their very large $P_\text{spont}$, which easily surpasses the charge threshold for state recognition. Thus, it is viable to \emph{trade} part of $P_\text{sw}$ for improved data retention and endurance.

\bibliographystyle{bibliography/IEEEtran}
\bibliography{bibliography/mybib,bibliography/addbib}

\end{document}